\documentclass[aps,pre,twocolumn]{revtex4-1}
\usepackage{hyperref}
\usepackage{graphicx, subfigure}
\usepackage{amsmath,amssymb}
\usepackage{float}
\usepackage{color}

\hypersetup{
  colorlinks=true,
  linkcolor=blue,
  citecolor=blue,
  linktoc=page,
  urlcolor=blue
}

\begin{document}

\title{Directed transport in quantum  star graphs}
\author{Jambul Yusupov}
\affiliation{Turin Polytechnic University in Tashkent, 17 Niyazov Street, 100095 Tashkent, Uzbekistan}
\author{Maxim Dolgushev}
\author{Alexander Blumen}
\author{Oliver M\"ulken}
\affiliation{
Physikalisches Institut, Universit\"at Freiburg,
Hermann-Herder-Stra{\ss}e 3, 79104 Freiburg, Germany}

\date{\today}

\begin{abstract}
We study the quantum dynamics of Gaussian wave packets on star graphs whose arms feature each a periodic potential and an external time-dependent
field. Assuming that the potentials and the field can be manipulated separately for each arm of the star, we show that it is possible to manipulate
the direction of the motion of a Gaussian wave packet through the bifurcation point by a suitable choice of the parameters of the external fields. In
doing so, one can achieve a transmission of the wave packet into the desired arm with nearly 70\% while also keeping the shape of the wave packet
approximately intact. Since a star graph is the simplest element of many other complex graphs, the obtained results can be considered as the first
step to wave packet manipulations on complex networks.
\end{abstract}

\maketitle

\section{Introduction}\label{sec:intro}

Recent experimental advances have led to very precise manipulations of atoms in optical lattices \cite{bloch2005} and of wave packets in waveguide
arrays \cite{szameit_2010, block_2014}. Fundamental effects such as Bloch oscillations for wave packets in tilted lattices have been experimentally
confirmed by several groups \cite{dekorsy_1995,dahan1996,haller_2010,lehtinen_2012}. Moreover, it has been shown
that coherent control over the wave packets is possible by varying the external field: Such variations can be discrete \cite{hartmann2004,breid_2007}
or continuous \cite{thommen2004,arlinghaus2011,creffield2011,hu2013}. In a two-dimensional lattice, one can achieve arbitrary displacements of the
wave packet by suitable modulation of the external field \cite{thommen2011}.

Also excitations (excitons) in idealized linear polymers, modeled by beads and springs, with an external field and at ultra-cold temperatures have
been shown to exhibit Bloch oscillations \cite{mulken2011}. Similar to the experiments mentioned above, one can manipulate the excitons' motion by
varying the external field \cite{thommen2002}.

However, all the experiments so far consider lattice-like underlying potentials. For more complex arrangements one encounters the situation of a
(quantum) graph \cite{kottos1999, gnutzmann2006}, where vertices are connected by arms. In order to be able to manipulate a wave packet on an
arbitrary graph, one first has to understand the behavior of wave packets at bifurcation points. These points are characteristic for vertices where
three (or more) arms meet. We call a graph with a single bifurcation a star graph. In the following, we will assume that each arm is modeled by a
one-dimensional periodic potential and that at the vertex there is continuity and current conservation, see below.

As we will show, it is possible to manipulate a wave packet by suitable changes of the external fields. Since complex graphs can be build iteratively
by joining star graphs, our results are a first step to precise manipulations of wave packets on these complex graphs. This paper is organized as
follows. In the next section we will recall the problem of a tilted lattice solved in a continuous model. Section \ref{sec:field_mod} presents the
study of the directed transport in a one dimensional lattice. In section \ref{sec:trans_sg} we treat such transport in driven quantum star graphs.
Finally, section \ref{sec:conclusion} presents concluding remarks.

\section{Bloch oscillations in a one-dimensional lattice}
\label{sec:bloch_1d}

\begin{figure}
\center
\includegraphics[trim=1.3cm 0cm 1.3cm 0cm, clip=true, width=8.5cm]{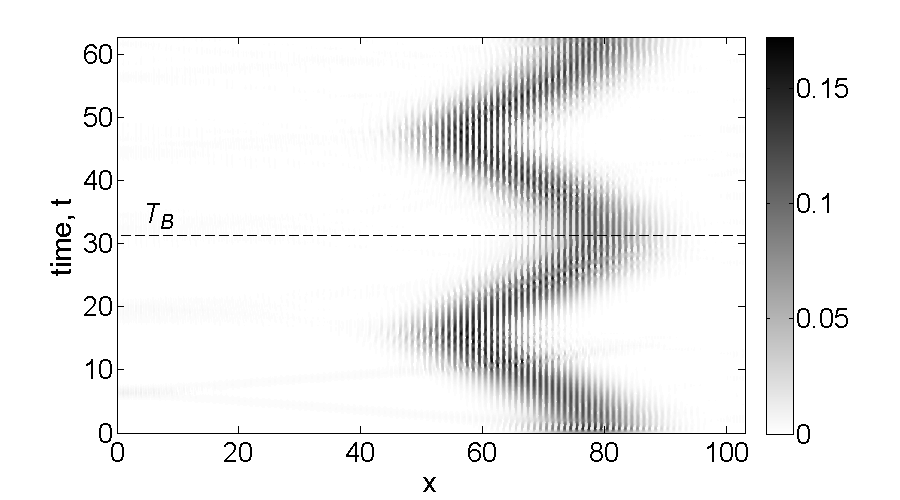}
\caption{Contour plot of the probability density. Bloch oscillations for $f=0.2, d=1$ and $V_0=16.7875$. One can see that the oscillation is bounded
within the interval of the length $\Lambda=\Delta/f=20$.} \label{figstat1d}
\end{figure}

We start by recalling the effect of Bloch oscillations in one-dimensional periodic potentials $V(x)$. The Hamiltonian is given (in units
$\hbar=2m=1$) by:
\begin{equation}\label{H0}
H=-\frac{\partial^2}{\partial x^2} +V(x) + fx\ ,
\end{equation}
where $f$ is the external field strength and $V(x+d)=V(x)$ is the periodic potential with lattice period $d$. This Hamiltonian describes, for
instance, an electron in a one-dimensional crystal in the presence of a constant electric field \cite{hartmann2004,gluck2002}. Such a Hamiltonian
generates time-periodic oscillations, the Bloch oscillations, of an initial Gaussian wave packet (GWP). These oscillations have a period inversely
proportional to the field strength, i.e., $T_B=2\pi \hbar/df$, and a well-defined amplitude $L=\Delta/f$, where $\Delta$ is the bandwidth. In the
following we choose the lattice periodic potential as a cosine potential of the form:
\begin{equation}\label{potential}
V(x)=V_0\cos\left(\frac{2\pi}{d}x\right)\ .
\end{equation}
Exemplarily, we show in Fig.~\ref{figstat1d} the Bloch oscillations of the solution of the Schr\"{o}dinger equation with $H$ and an initial GWP over
two Bloch periods. We chose the parameters such that our results are comparable to one of the discrete model discussed in Ref.~\cite{mulken2011}. We
note that the solution of Eq.~(\ref{H0}) shows the different bands of the Bloch band structure \cite{hartmann2004}, while in the discrete model only
the lowest band was considered. However, also for the continuous solution, the main fraction of the excitation is bounded to the lowest band, see the
weak contribution of the higher bands with larger velocity in Fig.~\ref{figstat1d}.

\section{Periodic potentials with field modulations}\label{sec:field_mod}

Now, modulations of the potential or of the external field allow to manipulate the motion of the GWP. In order to see this, we assume either a
time-dependent potential $V(x,t)$ or a time-dependent and spatially homogeneous external field $F(t)$. Then, the Schr\"odinger equation in the
one-dimensional case has the general form
\begin{equation}\label{H}
i \frac{\partial}{\partial t}\Psi(x,t)=-\frac{\partial^2}{\partial x^2}\Psi(x,t)+\left[V(x,t)+F(t)x\right]\Psi(x,t)\ .
\end{equation}

\begin{figure}
\includegraphics[trim=1.3cm 0cm 1.3cm 0cm, clip=true, width=8.5cm]{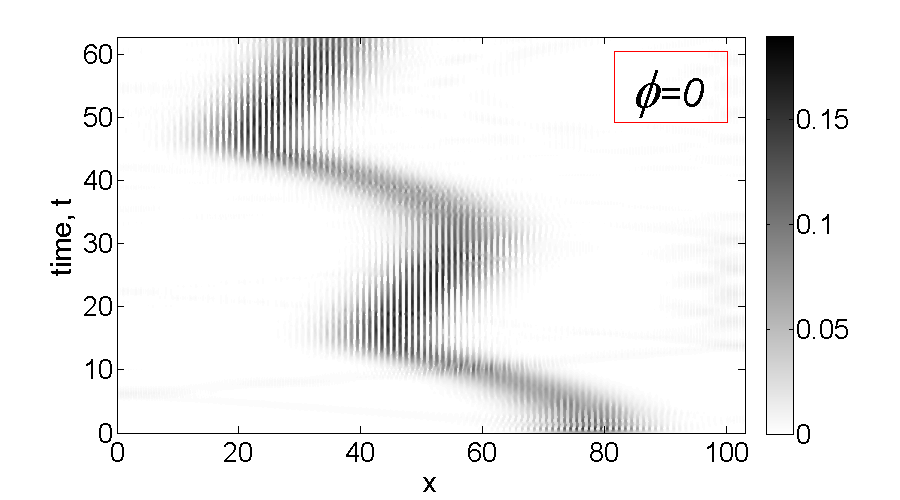}
\includegraphics[trim=1.3cm 0cm 1.3cm 0cm, clip=true, width=8.5cm]{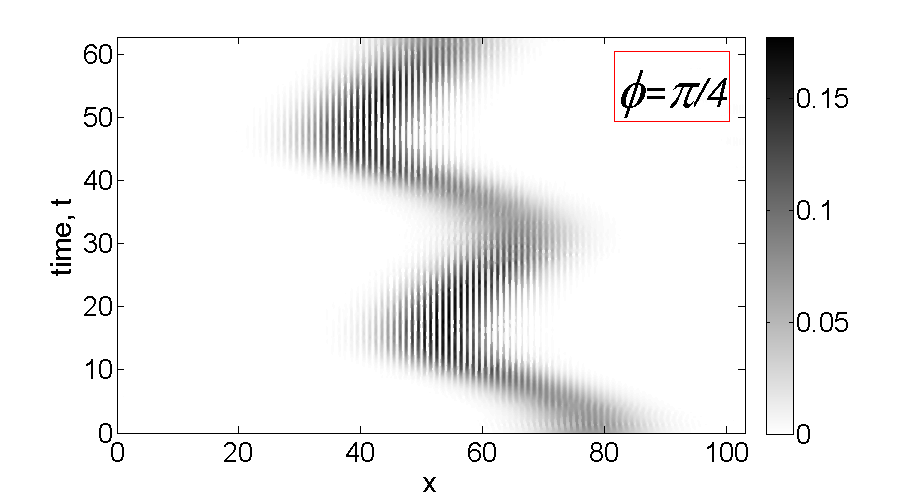}\\
\includegraphics[trim=1.3cm 0cm 1.3cm 0cm, clip=true, width=8.5cm]{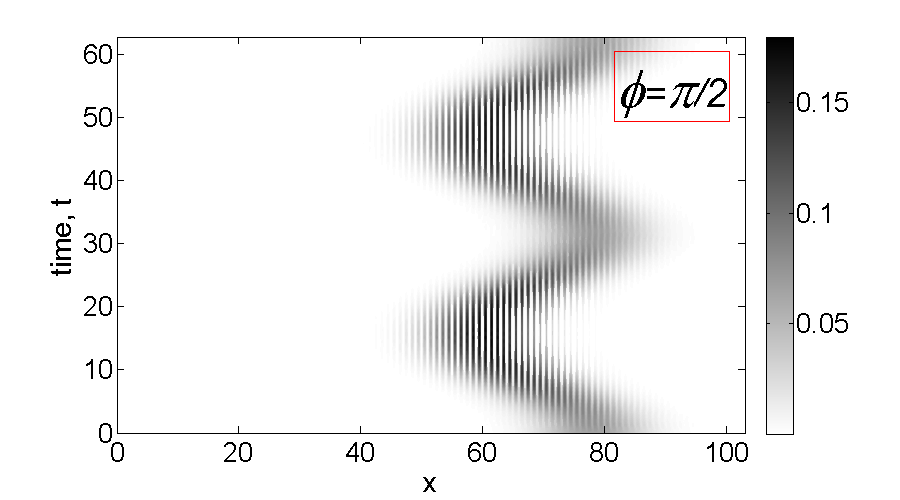}
\caption{Lattice potential amplitude modulation given by (\ref{V_pot}): Contour plot of probability density for $a=0.85$ and $\omega=df$. Each plot
corresponds to different values of $\phi=0,\pi/4$ and $\pi/2$.} \label{resvc}
\end{figure}

A simple sinusoidal change in the coupling strength is reflected by an oscillating potential amplitude leading to tunneling matrix elements of $H$
given by
\begin{equation}\label{V_pot}
V(x,t)=V_0[1-a\sin{(\omega t + \phi)}] \cos{\left(\frac{2\pi}{d}x\right)}\ ,
\end{equation}
where $a\in[0,1]$. Fig.~\ref{resvc} shows three cases of the dynamics of a GWP centered at $x_0=78$ with a standard deviation $\sigma=6$. Each panel
displays the motion for a constant external field of strength $f=0.2$ and an oscillating potential with parameters $d=1$, $a=0.85$, and $V_0=16.7875$,
but with different phases $\phi$. These parameters are such that the results match those of Ref.~\cite{mulken2011}, where the oscillating coupling
strength was achieved by varying distances between the centers of the (discrete) potential.

Experimentally it might be easier to modulate the external field than the potential. Thus, we consider in the following only the case of a
time-dependent field $F(t)=f\sin (\omega t + \phi)$ and a time-independent periodic potential $V(x)=V_0\cos[(2\pi/d) x]$. Note, that the phase $\phi$
is now also shifted to the field dependence. For periodically modulated external fields, it is known from the discrete model that a GWP with an
initial width $\sigma_0$ can broaden with time \cite{creffield2011}:
\begin{equation}\label{td_width}
\sigma(t)=\sigma_0\sqrt{1+  t^2 [\mathcal{J}_0(f/\omega)\cos{\left[(f/\omega)\cos{\phi}\right]}/\sigma_0^2]^2}\ ,
\end{equation}
where $\mathcal{J}_0(f/\omega)$ is the Bessel function of first kind. Therefore, whenever the Bessel function or the cosine function are zero, it is
possible to preserve the width of the initial GWP, leading to a non-dispersing wave packet for a GWP with $f=\pi/10, \omega=1/5$, and $\phi=0$. In
the following we will always use these parameters to preserve the width of the GWP (before the bifurcation point). Strictly, Eq.~(\ref{td_width}) is
only valid for the lowest band, while contributions from higher bands will in any case result in a broadening of the GWP.
We confirm the preservation of the width of the GWP in the lowest band by numerical fitting to a Gaussian with
parameterized width.

\begin{figure}
\includegraphics[trim=1.3cm 0cm 1.3cm 0cm, clip=true, width=8.5cm]{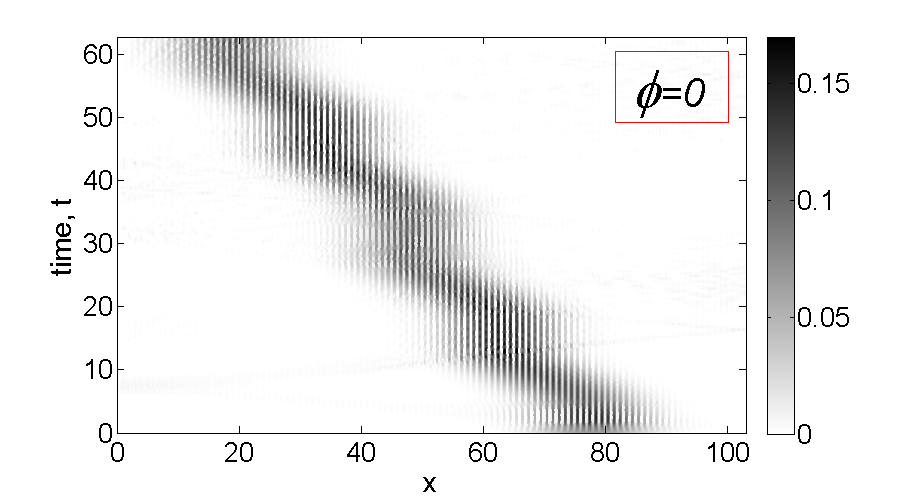}
\includegraphics[trim=1.3cm 0cm 1.3cm 0cm, clip=true, width=8.5cm]{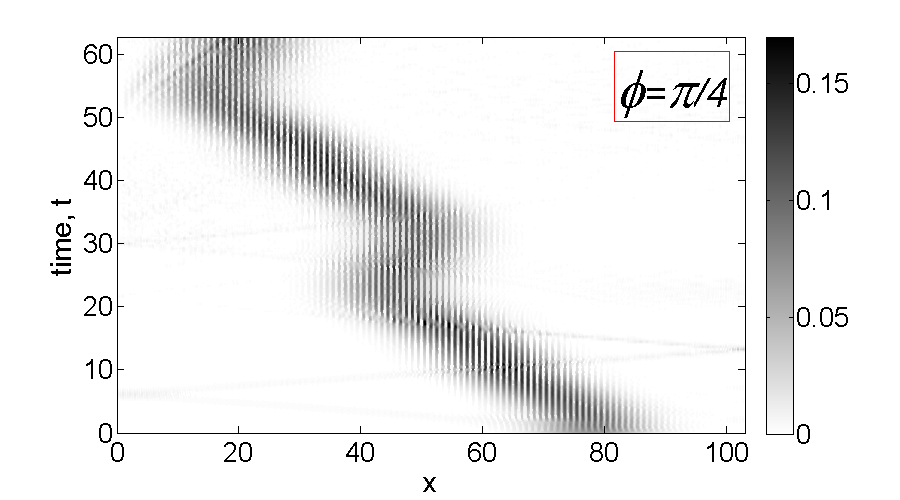}\\
\includegraphics[trim=1.3cm 0cm 1.3cm 0cm, clip=true, width=8.5cm]{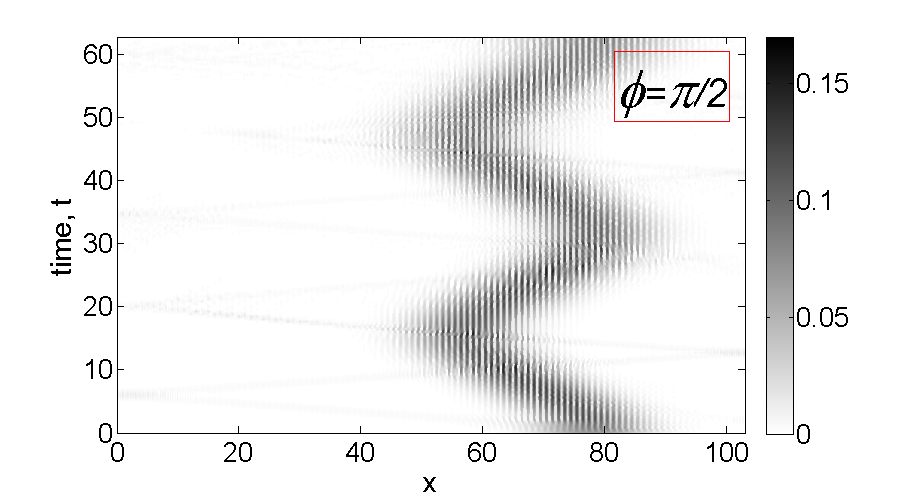}
\caption{ The external field modulation given by $F(t)=f\sin (\omega t + \phi)$: Contour plot of probability density for $f=\pi/10$ and $\omega=0.2$.
Each plot corresponds to different values of $\phi=0,\pi/4$ and $\pi/2$.} \label{resfc}
\end{figure}

Fig.~\ref{resfc} shows the dynamics of the same initial GWP as in Fig.~\ref{resvc} with parameters $f=\pi/10$, $d=1$, $\omega=0.2$, and $V_0=16.7875$,
for three different phases $\phi$. As for the case where the potential is time-dependent, one clearly observes the $\phi$-dependence of the dynamics:
While for $\phi=\pi/2$ there is no net displacement after one Bloch period, one can achieve a significant displacement for $\phi=\pi/4$ and $\pi=0$.
We note, that for time-dependent fields with a phase $\phi$, also the Bloch frequency depends on $\phi$.

\section{Directed transport in driven star graph}\label{sec:trans_sg}

In this section we solve the problem for a quantum star graph. Each of the arms is vested with a tilted lattice potential, and they are connected at
a common (central) vertex. We assign the coordinate $x$ to each arm, which indicates the position along the arm; $x$ takes the value 0 at the
common vertex.

Thus, we consider the time-dependent Schr\"{o}dinger equation for a star graph with $N$ arms ($\hbar=2m=1$):
\begin{equation}\label{sge}
i \frac{\partial}{\partial t}\Psi_j(x,t)=H_j\Psi_j(x,t),
\end{equation}
\begin{flushright}
\ $\forall t,0\leq x \leq L_j, \ \ j=1,...,N$\ \ \ \ \ \ \ \ \ \ \
\end{flushright}
where
$$
H_j=-\frac{\partial^2}{\partial x^2}+\left[V_0\cos\left(\frac{2\pi}{d}x\right)+F_j(t)x\right].
$$
As mentioned in the previous section, the time-dependence of the field is given by $F_j(t)=f_j\sin (\omega t + \phi_j)$, being in this case also
arm-dependent, and the parameters $f_j$ and $\phi_j$ are used to manipulate the GWP transition through the central vertex. The imposed conditions
\begin{equation}\label{sgbc}
\left\{
\begin{array}{l}
  \Psi_1(L_1,t)=\Psi_2(L_2,t)=...=\Psi_N(L_N,t)=0\ , \\
  \Psi_1(0,t)=\Psi_2(0,t)=...=\Psi_N(0,t)\ , \\
  \sum\limits_{j=1}^N{\frac{\partial}{\partial x}\Psi_j(x,t)|_{x=0}}=0\ . \\
\end{array}
\right.
\end{equation}
imply Dirichlet boundary conditions at the non-connected edges and continuity and current conservation at the vertex.

The solution of Eqs.~(\ref{sge})-(\ref{sgbc}) can be written in terms of the complete set of eigenfunctions $\psi_{j,n}(x)$ of the potential-free
star graph:
\begin{equation}\label{sgsol}
\Psi_j(x,t)=\sum\limits_n{C_n(t)\psi_{j,n}(x)}, \ \ j=1,\dots,N.
\end{equation}
where time-dependent coefficients $C_n(t)$ are to be found.

For the star graph with $N$ arms the eigenfunctions $\psi_{j,n}(x)$ of the stationary Schr\"odinger equation (in units $\hbar=2m=1$):
\begin{equation}\label{fsgea}
-\frac{d^2}{dx^2}\psi_j(x)=k^2\psi_j(x),\ \ 0\leq x \leq L_j, \ \
j=1,...,N,
\end{equation}
with the same boundary conditions as in Eq.~(\ref{sgbc}) have the form
\begin{equation}\label{fsgwf}
\psi_{j,n}(x)=\frac{B_n}{\sin{(k_n L_j)}} \sin{[k_n (L_j-x)]},
\end{equation}
where
\begin{equation}\label{fsgnc}
B_n=\left[\sum\limits_j{[L_j+\sin{(2 k_n L_j)}]\sin^{-2}{(k_n L_j)}}/2\right]^{-1/2}
\end{equation}
are the normalization coefficients, and the $k_n$ fulfill the equation:
\begin{equation}\label{fsges}
\sum\limits^N_{j=1}{\cot(k_n L_j)}=0.
\end{equation}
Eq.~(\ref{fsges}) can be solved numerically. In order to avoid non-generic degeneracies, the lengths of arms $L_j$ are chosen to be rationally
independent \cite{kottos1999}.

Inserting the expansion~(\ref{sgsol}) into Eq.~(\ref{sge}) and taking into account the orthonormality of the eigenfunctions,
\begin{equation}\label{sgnc}
\sum\limits_{j=1}^N{\int\limits_0^{L_j}{\psi_{j,m}^*\psi_{j,n} dx}}=\delta_{mn},
\end{equation}
we get the system of ODE with respect to the coefficients:
\begin{equation}\label{sgode}
i\dot{C}_n=k_n^2 C_n + \sum\limits_m{M_{nm}C_m}.
\end{equation}
The matrix $M$ can be written in terms of the contributions arising from the lattice potential ($I^{(V)}$) and the driving external field
($I^{(F)}$):
$$
M=I^{(V)}+I^{(F)},
$$
the elements of which are given in Appendix \ref{app:int}.

As the system of ODE (\ref{sgode}) has an infinite number of equations, solving it numerically requires to limit this number. Then, to preserve the
required accuracy one has to control the norm conservation condition
\begin{equation}\label{sgcc}
\sum\limits_n{|C_n(t)|^2}=1, \ \ \forall t.
\end{equation}

\begin{widetext}

\begin{figure}[H]
\center
\includegraphics[trim=1.4cm 0cm 0.8cm 0.8cm, clip=true, width=17cm]{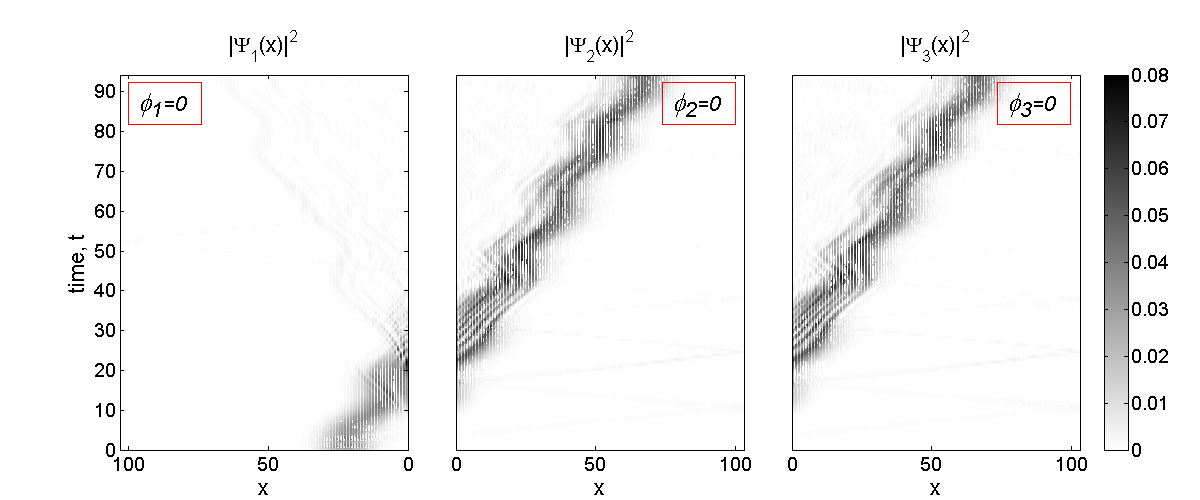}
\caption{Contour plot of the probability density for the three arm star graph, where driving external field is given by $F_j(t)=f_j\sin{(\omega
t+\phi_j)}$ with $f_1=-f_2=-f_3=\pi/10, \omega=0.2$. The three columns correspond to the arms of the star graph. The $x$-coordinate of the first
arm is reversed.} \label{gwp3bsg0}
\end{figure}

\end{widetext}

\begin{figure}
\center
\includegraphics[width=8.5cm]{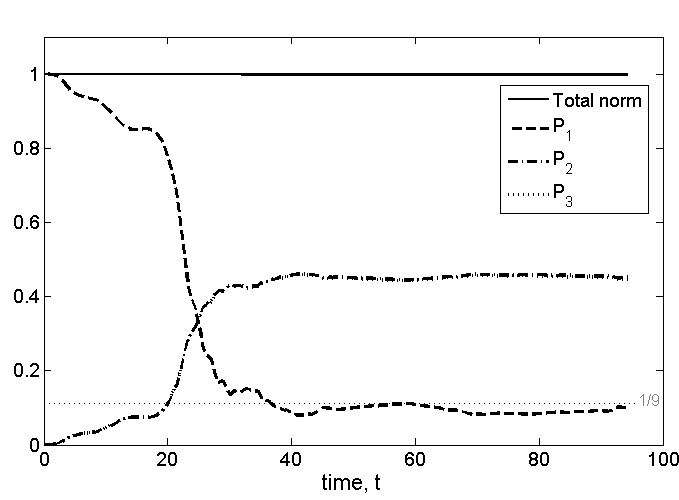}\\
\caption{The time-dependence of the partial norms corresponding to Fig.~\ref{gwp3bsg0}.} \label{gwp3bsg0_pn}
\end{figure}

We now turn to the results for a GWP on a star graph. For clarity, we restrict ourselves to a star graph with three arms. Fig.~\ref{gwp3bsg0} shows
the dynamics of a GWP initially located on the first arm, at $x_0=22$, see the leftmost panel in Fig.~\ref{gwp3bsg0}. In this first example, the
external field strengths $F_j(t)=f_j\sin (\omega t + \phi_j)$ are of the same magnitude $f$ for all the arms and chosen to be $f_1=-f_2=-f_3$, such
that the field points globally in one direction. As for the case of Bloch oscillations on a single arm, the GWP moves towards the vertex (at position
$0$). At the vertex the GWP bifurcates and is partly transmitted to the other two arms and partly reflected back into the initial arm. As the driving
external field for the second and third arms has the same parameters, the dynamics of the split GWP is identical for these two arms. This can also
be seen in Fig.~\ref{gwp3bsg0_pn}, where the time-dependence of the partial norms
$$
P_j(t)=\int\limits_0^{L_j}{|\Psi_j(x,t)|^2dx}, \ \ j=1,2,..,N
$$
are shown. We have confirmed that the total norm is $\sum_{j=1}^N{P_j(t)}=1$. On the basis of previous results for quantum graphs  with the Neumann
boundary condition \cite{kottos1999}, we can calculate the fraction of the GWP which is reflected back into the initial arm as being $1/9$. The other
fraction of $8/9$ of the initial GWP is split into the two identical parts. Due to the reflection and transmission at the vertex the shape of the GWP
in the two arms is (slightly) distorted, but with a width which is still roughly preserved over time.

\begin{widetext}

\begin{figure}[H]
\center
\includegraphics[trim=1.4cm 0cm 0.8cm 0.8cm, clip=true, width=17cm]{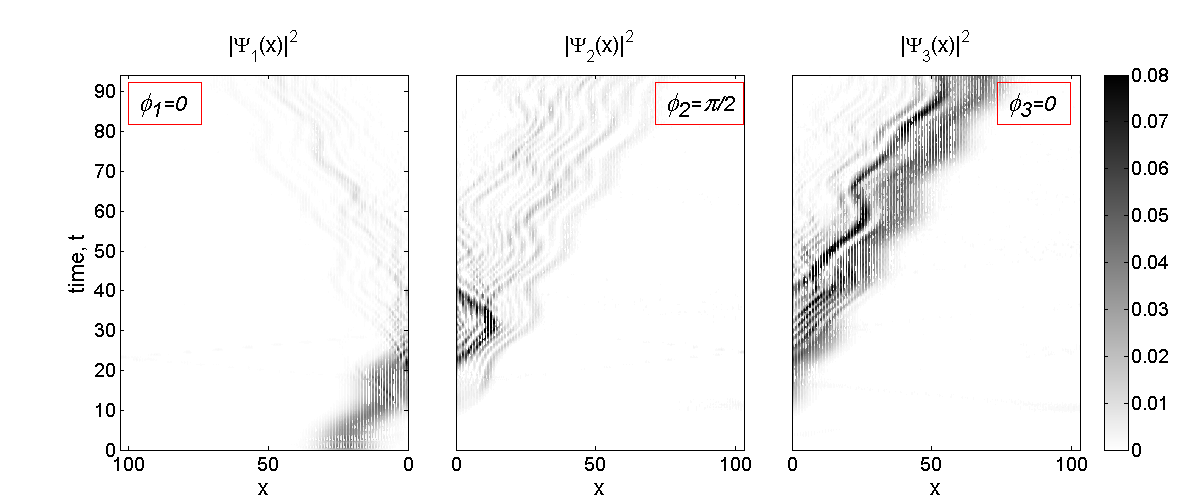}
\caption{Contour plot of the probability density for the three arm star graph, where driving external field is given by $F_j(t)=f_j\sin{(\omega
t+\phi_j)}$ with $f_1=-f_2=-f_3=\pi/10, \omega=0.2$. The three columns correspond to the arms of the star graph. The $x$-coordinate of the first
arm is reversed.} \label{gwp3bsg90}
\end{figure}

\end{widetext}

\begin{figure}
\center
\includegraphics[width=8.5cm]{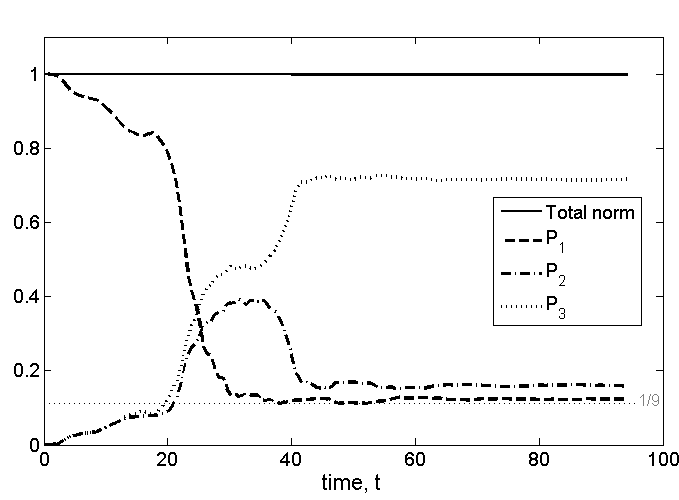}
\caption{The time-dependence of the partial norms corresponding to Fig.~\ref{gwp3bsg90}.} \label{gwp3bsg90_pn}
\end{figure}

The situation changes when we change the phase of the second arm to $\phi_2=\pi/2$, see Fig.~\ref{gwp3bsg90}. While the transmission into the third
arm persists, although the wave packet becomes more distorted, the transmission in the second arm is nearly prohibited. This effect can be
quantified by calculating the time-dependence of the partial norms, which are shown in Fig.~\ref{gwp3bsg90_pn}. After a transient time until most
reflection and transmission effects have taken place, the partial norm in arm three saturated around a value of 70\%, while the partial norms of the
other two arms saturate around values between 10 and 20 percent. Even though the wave packet in the third arm gets distorted it still retains a
rather constant width. Therefore, by manipulating the phase of the external field in one of the arms, one is able to ``block'' the transmission into
this arm and consequently can manipulate the direction of the initial GWP.

\begin{figure}
\center
\includegraphics[width=8.5cm]{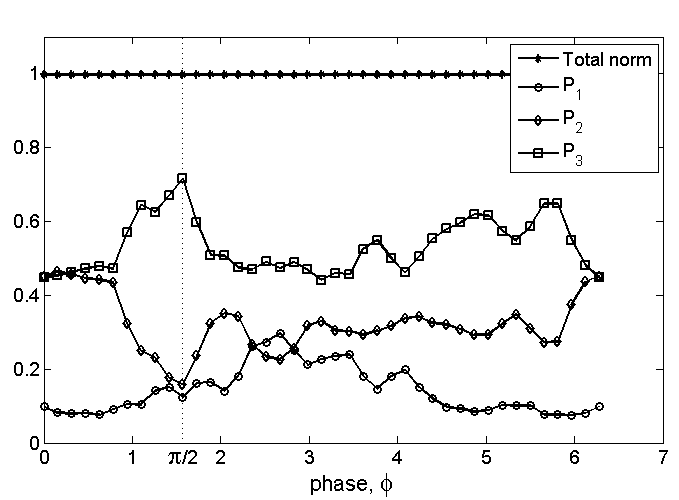}\\
\caption{The phase-dependence of the partial norms at time $t=94.25$. The external field and initial parameters correspond to Figs.~\ref{gwp3bsg0}
and \ref{gwp3bsg90}.} \label{phase_pn}
\end{figure}

For our setup of initial parameters, the choice $\phi_2=\pi/2$ turns out to be the most efficient one to realize the idea of ``blocking'' the
wave packet propagation to the second arm. To show this, we calculated the partial norms for the three arms depending on the phase $\phi_2$.
Fig.~\ref{phase_pn} shows the value of the partial norms at times where the norms have saturated, i.e., we have chose that time at which we stop our
numerical calculations, $t=94.25$. At the value of $\phi_2=\pi/2$, we observe the largest difference in the partial norms.

\section{Conclusion}\label{sec:conclusion}

In this paper, we have investigated the possibilities of directing a Gaussian wave packet on a star graph with three arms. This can be viewed as a
paradigmatic building block for complex networks, which can be composed by iteratively joining star graphs. Assuming that each arm of the graph is
equipped with a periodic potential and additionally (independent) possibly time-dependent external fields act on each arm, our numerical results show
that it is possible to direct a wave packet with high probability from one arm into one of the other arms while blocking the transmission into the
third arm. This effect is best achieved by assuming time-periodic external fields with independent phase shifts. The phase shifts allow for a
parametric tuning of the ``blocking'' effect. Our results further indicate that the shape of the wave packets remains almost intact. We believe that
such a scenario can be realized by state-of-the-art experiments with, say, ultra-cold atoms in optical lattices. One can also imagine that after
transmission of the wave packet into one arm, the shape of the wave packet can be re-established, see, e.g., \cite{uberna1999phase}. The fact that
there is not perfect transfer into the desired arm can be explained by fundamental reflection and transmission rules at the bifurcation point, see
also \cite{kottos1999}. This study being a proof of concept, we will extend our analysis to more complex situations where several star graphs are
joined together forming more complex networks, such as, e.g., T-fractals \cite{agliari2008} or dendrimers \cite{mulken2006}.

\begin{acknowledgments}
We appreciate the clarifying input from Marcel Mudirch. Further, we thank the Deutscher Akademischer Austauschdienst (DAAD Grant No. 56266206 and
project no.\ 40018). We further acknowledge support from the Deutsche Forschungsgemeinschaft (DFG Grant No. MU2925/1-1), from  the Fonds der
Chemischen Industrie, and from the Marie Curie International Research Staff Exchange Science Fellowship within the 7th European Community Framework
Program SPIDER (Grant No. PIRSES-GA-2011-295302).
\end{acknowledgments}

\onecolumngrid

\appendix

\section{Calculation of matrices $I^{(V)}$ and $I^{(F)}$}\label{app:int}

The elements of the matrices $I^{(V)}$ and $I^{(F)}$ have integral forms, which can be solved analytically. Denoting the frequency of the periodical
lattice potential by $\omega_d=2\pi/d$ one gets the following:

\begin{eqnarray}
I_{nm}^{(V)}&=&V_0\sum\limits_{j=1}^N{\int\limits_0^{L_j}{\psi_{j,m}^* \cos{(\omega_d x)} \psi_{j,n} dx}} \\ \nonumber &=&
V_0\sum\limits_{j=1}^N{\frac{B_nB_m}{4\sin{(k_nL_j)}\sin{(k_mL_j)}}}\left[ \frac{\sin{(\omega_d L_j)}+\sin{(k_n-k_m)L_j}}{\omega_d+k_n-k_m} +
\frac{\sin{(\omega_d L_j)}-\sin{(k_n-k_m)L_j}}{\omega_d-k_n+k_m}   \right.\\ \nonumber & & \hskip 4.35cm -\left. \frac{\sin{(\omega_d
L_j)}+\sin{(k_n+k_m)L_j}}{\omega_d+k_n+k_m} - \frac{\sin{(\omega_d L_j)}-\sin{(k_n+k_m)L_j}}{\omega_d-k_n-k_m}\right],
\end{eqnarray}

\begin{equation}
I_{nm}^{(F)}=\sum\limits_{j=1}^N{F_j(t)\int\limits_0^{L_j}{\psi_{j,m}^* x \psi_{j,n} dx}}=\sum\limits_{j=1}^N{\frac{F_j(t) B_nB_m}{\sin{(k_n
L_j)}\sin{(k_m L_j)}}A_{nm}},
\end{equation}
where the diagonal elements of matrix $A$ (i.e. for $m=n$) are:
$$
A_{nn}=\frac{L_j^2}{2} - \frac{1-\cos{(2k_nL_j)}}{4k_n^2}
$$
and the off-diagonal elements ($m\neq n$) are:
$$
A_{nm}=\frac{1-\cos{2(k_n-k_m)L_j}}{(k_n-k_m)^2} - \frac{1-\cos{2(k_n+k_m)L_j}}{(k_n+k_m)^2}.
$$

\twocolumngrid

\bibliography{bibfile}

\end{document}